# Observing Jupiter's polar stratospheric haze with HST/STIS.
## An HST White Paper

Denis GRODENT, Bertrand BONFOND
*Laboratory for Planetary and Atmospheric Physics*
*Université de Liège, Belgium*

Jonathan NICHOLS
*Physics and Astronomy Department*
*University of Leicester, UK*

d.grodent@ulg.ac.be

11 August 2015

## 1. ABSTRACT

The purpose of this HST white paper is to demonstrate that it is possible to monitor Jupiter's polar haze with HST/STIS without breaking the ground screening limit for bright objects. This demonstration rests on a thorough simulation of STIS output from an existing image obtained with HST/WFPC2. It is shown that the STIS NUV-MAMA + F25CIII filter assembly provides a count rate per pixel ~11 times smaller than that obtained for one pixel of WFPC2 WF3 CCD + F218W corresponding filter. This ratio is sufficiently large to cope with the bright solar light scattered by Jupiter's atmosphere, which was a lesser concern for WFPC2 CCD safety. These STIS images would provide unprecedented spatial and temporal resolution observations of small-scale stratospheric aerosol structures, possibly associated with Jupiter's complex FUV aurora.

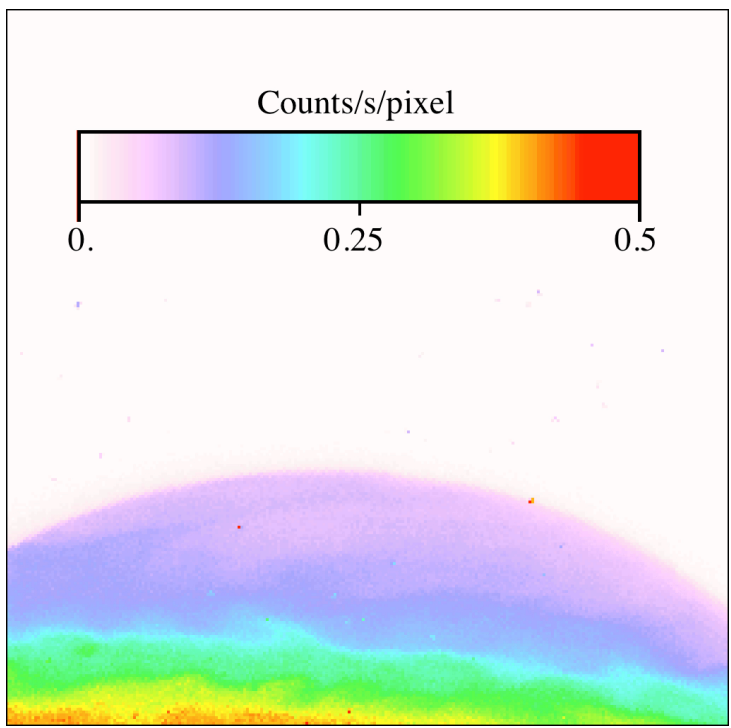

**Figure 1:** Simulated STIS NUV-MAMA image of Jupiter's northern polar region obtained by scaling an existing WFPC2 (WF3) image. The color table provides adjusted count rates in counts per sec per STIS pixel. The image illustrates the large absorption of MUV light by stratospheric haze in the polar region that makes it possible to safely observe this part of Jupiter's MUV disk. It also shows the stratospheric haze structures (hues of violet) that might be associated with auroral precipitation. In this simulated image, the global count rate is 61,121 counts.s$^{-1}$; that is 3.3 times lower than the screening global limit. (Bright dots are traces of WFPC2 cosmic rays).

## 2. INTRODUCTION

The 200-300 nm MUV range is particularly useful for studying the giant planets' stratosphere. In that range, isolated layers of polar atmospheric haze are absorbing solar UV light that would normally be efficiently Rayleigh scattered towards Earth by the neutral atmosphere. Therefore, it provides a unique tool for remotely characterizing the morphology of these polar stratospheric structures and to trace their motions. Jovian stratospheric haze is thought to result from nucleation and coagulation of heavy hydrocarbons such as benzene ($C_6H_6$) and polycyclic aromatic hydrocarbons (PAHs) (Friedson et al., 2002). Both are efficiently absorbing light in the FUV-MUV wavelength range (e.g. Verstraete et al., 1992).

## 3. STIS BRIGHTNESS RESTRICTION

Observations of the polar haze stratospheric structures were obtained with the WFPC2 instrument using the F218W, F255W or FQCH4N15 filters (Vincent et al., 2000), and later, with the ISS/NAC instrument on board the Cassini spacecraft (Porco et al., 2003). Short time (<1 sec) STIS observations would significantly enhance the quality of the data and provide a much more detailed description of the dark spots morphology that would make it possible to link them with specific components of Jupiter's aurora (Vincent et al., 2000; Porco et al., 2003; Gérard et al., 1995). However, the use of STIS for bright planetary UV targets is severely restricted/prohibited by the absolute MAMA count rate limits (ISR STIS 96-028). According to the STIS Instrument Handbook (IHB) (v14.0, 2015, p162) "Jupiter and Saturn are much too bright to be observed with most STIS NUV-MAMA imaging modes". Assuming that Jupiter's MUV emission disk is a relatively constant extended source, brightness restrictions set a ground screening global limit of 200,000 $counts.s^{-1}$ and a local limit of 100 $counts.s^{-1}.pix^{-1}$.

In the following, we demonstrate that when the Jovian disk is properly positioned in the field of view of STIS NUV-MAMA + F25CIII filter, the global count rate is comfortably below the screening limit. This situation is somewhat similar to the case of the FUV auroral images observed with STIS FUV-MAMA and for which several thousands of observations have been obtained without causing any damage to the instrument.

## 4. PROCEDURE

The method that we have used is based on the thorough scaling of an existing WFPC2 MUV (typical) image of Jupiter in the F218W filter. This scaling realistically simulates what would be observed with STIS NUV-MAMA + F25CIII filter in the same viewing conditions.

### 4.1. Throughputs
The WFPC2 to STIS scaling factor mainly depends on the total system spectral throughputs of these instruments (optical telescope assembly + detector + filter). For the sake of clarity, we used the STIS F25CIII NUV-MAMA Integrated System Throughput and Redleak, displayed in figure 14.55 of STIS IHB v14.0 and the WFPC2 F218W Total System Throughput QT, displayed in appendix A.1.3 of the WFPC2 IHB v10.0.

### 4.2. Emission source
The scaling factor also depends on the spectral radiance, or specific intensity, of the emission source. In our case, the emission is the result of the Rayleigh scattering of the solar light by Jupiter's neutral atmosphere and its attenuation by stratospheric hydrocarbons. Accordingly, the emission spectrum may be modelled by convolving the solar spectrum by Jupiter's spectral reflectance (single scattering albedo). We arbitrarily selected the recent (10 July 2015) solar spectral irradiance measured by the SORCE/SOLSTICE instrument and distributed by LASP at the University of Colorado (http://lasp.colorado.edu/home/sorce/data/ssi-data/). The MUV irradiance spectrum is given in units of $W.m^{-2}.nm^{-1}$ in the 180-310 nm range with a 1-nm resolution. We extended this range to larger wavelengths, up to 500 nm , with the spectrum provided by the SORCE/SIM instrument, in order to account for the important "red leak" of the STIS F25CIII NUV-MAMA assembly.

4.3. Reflectance

For the Jovian reflectance, we combined the center-of-disk I/F FUV (140-220 nm) and MUV (220-320 nm) hand-digitized spectra provided by Wagener et al. (1985). For simplicity, we assumed isotropic reflectivity (Lambertian approximation) at zero phase angle. Finally, we scaled down the solar irradiance at Earth orbit (1 AU) by a factor of 1/27 (the inverse of Jupiter's square distance to the Sun, 5.2 AU) and we adjusted the units so that the resulting Jovian spectral radiance is given in $erg.s^{-1}.cm^{-2}.Å^{-1}.arcsec^{-2}$.

## 5. WFPC2 AND STIS COUNT RATES

The count rates were estimated by using the equations given in the WFPC2 and STIS IHBs for extended sources imaging. For WFPC2, we considered equation (6.13), appropriate for an emission line observed with one of the Wide Field Cameras and integrated it over the MUV bandwidth. This provides the estimated count rate in electrons per sec per pixel, which is then divided by the CCD gain (G=7) to obtain the count rate in $DN.s^{-1}.pix^{-1}$. For STIS, we used the equation given in section 6.2.2 of STIS IHB for a diffuse source (for $N_{pix}$ and $G = 1$). With these equations, throughputs and Jovian emission spectrum described above, we obtain a count rate of 3.39 $DN.s^{-1}.pix^{-1}$ for WFPC2 (WF3) + F218W and 0.40 $counts.s^{-1}.pix^{-1}$ for STIS NUV-MAMA + F25CIII. These numbers are giving rise to a WFPC2-to-STIS scaling factor of ~10, meaning that when observing the same target, one 0.1 arcsec wide WFPC2 pixel maps to a 10 times fainter 0.025 arcsec STIS pixel.

We note that our WFPC2 (WF3) count rate (3.39 $DN.s^{-1}.pix^{-1}$) is close to the value derived by Karkoschka (1998) (4.8 $DN.s^{-1}.pix^{-1}$). The discrepancy is essentially owing to updated throughput, spectral coverage and resolution. The estimated STIS count rate (0.4 $counts.s^{-1}.pix^{-1}$) is far below the screening local limit of 200 $counts.s^{-1}.pix^{-1}$. We used the Exposure Time Calculator (ETC ver. 23.2) provided by STScI to crosscheck our results for STIS. We assumed an extended source with a diameter of 16 arcsec, representing the ~1/8 portion of Jupiter's disk appearing in STIS FOV (see section below), and uploaded the spectral distribution file containing the Jovian spectral radiance described above. With this tool, the expected STIS count rate is 0.42 $counts.s^{-1}.pix^{-1}$ (ETC Request ID: STIS.im.736453), in very close agreement with the value derived above (0.40).

At this point it should be emphasized that these count rates do not take into account the important darkening of the polar regions. Therefore, they provide upper limits to the actual count rates. A more realistic estimation may be obtained by scaling and positioning an existing WFPC2 image in STIS FOV, as described below.

## 6. IMAGE SIMULATION

We considered one typical MUV image of Jupiter's disk obtained with WFPC2 (WF3) + F218W in the frame of HST-GO program 6509 (K. Rages, PI). We selected image *u3ay0205m*, as it conveniently displays Jupiter's full disk. The image was acquired for 260 sec on 25 June 1997 at 11:16UT. It was extracted from the retrieved *.c0m* science file and then geometrically corrected (Gilmozzi et al., 1995). In this image, the actual count rate averaged over the full Jovian disk is 4.2 $DN.s^{-1}.pix^{-1}$, again very close to our estimated count rate (3.39). In order to be consistent with the actual data counts, we increased the WFPC2 throughput by a factor of 1.24, so that we obtain a similar count rate of 4.2 $DN.s^{-1}.pix^{-1}$ from the estimation procedure described above. This is slightly changing the WFPC2 to STIS scaling factor, which is now 10.6. We enlarged the WFPC2 image by a factor of 4 to account for the ~4 times higher spatial resolution of STIS and divided the resulting image by the 10.6 corrected scaling factor. We then cropped the image to the actual FOV of STIS for different positions of Jupiter. The resulting image is displayed in **Figure 1**.

## 7. RESULTS

In the worst-case scenario, Jupiter is approximately centred in the FOV. Accordingly, the FOV is almost entirely filled with the brightest region of Jupiter's disk, while missing the attenuated polar regions. In this case, the total count rate is 524,612 counts.s$^{-1}$; 2.6 times larger than the screening global limit (200,000). In a more realistic --and useful case, Jupiter's disk is off-centered, so that the image displays approximately 1/8 of the Jovian disk and focuses on one of the darker polar regions. This viewing geometry is similar to that commonly used for observing the Jovian aurora with STIS FUV-MAMA. In this case, **the global count rate drops to 61,121 counts.s$^{-1}$; that is 3.3 times lower than the screening global limit**. If one multiplies the ETC pixel count rate (4.2) by the corresponding number of visible Jovian disk pixels (337,715) then we get a global count rate of 141,840 counts.s$^{-1}$, still below the limit. This multiplicative factor of 2.3 clearly illustrates the overestimation of the global count rate if one neglects the darkening of the polar regions.